%% file: paper_la.tex
\documentclass[11pt]{article}
\usepackage{graphicx}

\usepackage{epsfig}
\usepackage{dcolumn}
\usepackage{amsmath}
\usepackage{multirow}
\usepackage{cite}

\newcommand{\BABARPubYear}    {08}
\newcommand{\BABARConfNumber} {008}
\newcommand{\SLACPubNumber} {13367}

\input pubboard/babarsym
 
\def\effkpi{20.6}                        
\def\effkspi0{11.7}                        
\def\effkpi0{13.7}                         
\def\effkspi{18.8}

\def\kszkpbr{4.55}
\def\kszkpbrstat{0.11}
\def\kszkpbrsyst{0.16}

\def\kszkszbr{5.01}
\def\kszkszbrstat{0.40}
\def\kszkszbrsyst{0.37}

\def\kspkpzbr{5.05}
\def\kspkpzbrstat{0.22}
\def\kspkpzbrsyst{0.27}

\def\kspkspbr{4.56}
\def\kspkspbrstat{0.20}
\def\kspkspbrsyst{0.17}

\def\kszcombr{4.58}
\def\kszcomstat{0.10}
\def\kszcomsyst{0.16}

\def\kspcombr{4.73}
\def\kspcomstat{0.15}
\def\kspcomsyst{0.17}

\def\isoval{0.029}
\def\isostat{0.019}
\def\isosyst{0.016}
\def\prodratio{0.018}

\def\isolow{-0.021}
\def\isohi{0.079}

\def\acpval{-0.009}
\def\acpstat{0.017}
\def\acpsyst{0.011}

\def\acplolim{-0.043}
\def\acphilim{0.025}

\def\ifb{\;\mbox{fb}^{-1}}
\def\lumitotal{347\ifb}

\def\numBB{383 million}

\def\GeV{\;\mbox{GeV}}

\def\GeVcc{\;\mbox{GeV}/c^2}

\def\de   {\Delta E}
\def\dz   {\Delta z}
\def\mes  {m_{\mbox{\scriptsize ES}} }
\def\bkg  {B \to K^{*}\gamma}
\def\bdpi  {B \to D\pi}
\def\bkgneut  {B^0 \to K^{*0}\gamma}
\def\bkpg  {B^+ \to K^{*+}\gamma}

\def\bsg    {b\to s\gamma}

\def\kszkp  {K^{*0} \rightarrow K^+\pi^-}
\def\kspkpz {K^{*+} \rightarrow K^+\pi^0}
\def\kspksp {K^{*+} \rightarrow K_S\pi^+}
\def\kszksz {K^{*0} \rightarrow K_S\pi^0}

\def\incbsg  {B\to X_s\gamma}

\def\acp {\ensuremath{\mathcal{A}}}
\def\aexcl {\ensuremath {  \acplolim <\acp(\bkg) < \acphilim }}
\def\isoexcl{\ensuremath {\isolow < \Delta_{0-} < \isohi }}

\def\isoas{\Delta_{0-}}

\def\B       {\ensuremath{B}\xspace}
\def\Bbar    {\kern 0.18em\overline{\kern -0.18em B}{}\xspace}

\def\BB      {\ensuremath{B\Bbar}\xspace} 

\def\bkg    {\ensuremath {\B \to \Kstar \gamma}}

\def\bkpg    {\ensuremath {\Bp \to \Kstarp \gamma}}

\def\Kz    {\ensuremath{K^{0}}\xspace}

\def\CP                {\ensuremath{C\!P}\xspace}

\long\def\inst#1{\par\nobreak\kern 4pt\nobreak
    {\it #1}\par\vskip 10pt plus 3pt minus 3pt}

\begin{document}
{\pagestyle{empty}

\begin{flushright}
\babar-CONF-\BABARPubYear/\BABARConfNumber \\
SLAC-PUB-\SLACPubNumber \\
\end{flushright}

\par\vskip 5cm

\begin{center}
\Large \bf Measurement of Branching Fractions and $CP$ and Isospin Asymmetries in $B\to K^*\gamma$
\end{center}
\bigskip

\begin{center}
\large The \babar\ Collaboration\\
\mbox{ }\\
\today
\end{center}
\bigskip \bigskip

\begin{center}
\large \bf Abstract
\end{center}
We present a preliminary analysis of the decays $\bkgneut$ and $\bkpg$ using a sample of \numBB~\BB\ events collected with the \babar\ detector at the PEP-II asymmetric energy $B$ factory.  We measure the branching fractions ${\cal B}(\bkgneut) =  (\kszcombr \pm \kszcomstat \pm \kszcomsyst)\times 10^{-5}$  and ${\cal B}(\bkpg) =  (\kspcombr \pm \kspcomstat \pm \kspcomsyst) \times 10^{-5}$.  We measure the direct \CP asymmetry to be \aexcl~and the isospin asymmetry to be \hbox{\isoexcl}, where the limits are determined at the 90\% confidence interval and include both the statistical and systematic uncertainties.
\vfill
\begin{center}

Submitted to the 33$^{\rm rd}$ International Conference on High-Energy Physics, ICHEP 08,\\
30 July---5 August 2008, Philadelphia, Pennsylvania.

\end{center}

\vspace{1.0cm}
\begin{center}
{\em Stanford Linear Accelerator Center, Stanford University, 
Stanford, CA 94309} \\ \vspace{0.1cm}\hrule\vspace{0.1cm}
Work supported in part by Department of Energy contract DE-AC02-76SF00515.
\end{center}

\newpage
} 

\input pubboard/authors_ICHEP2008

\section{INTRODUCTION}
\label{sec:Introduction}

In the Standard Model (SM), the decays \mbox{\bkg}~\cite{Kstref} proceed dominantly through one-loop \mbox{$\bsg$} electromagnetic penguin transitions.  Extensions of the SM predict new high-mass particles that can exist in the loop and alter the SM prediction of the branching fractions.  The theoretical predictions of the decay rates~\cite{Ali:2001ez, Bosch:2001gv, Beneke:2001at,  Matsumori:2005ds} for \mbox{\bkg} suffer from large hadronic uncertainties, and previous measurements of the branching fractions (Table~\ref{tab:prevMeasurements}) are more precise than SM estimates.  The theoretical estimates and experimental measurements of the branching fractions are in reasonable agreement.

Experimental and theoretical uncertainties are much reduced when considering the $CP$ and isospin asymmetries~\cite{CC}, which are defined by:

\begin{equation}
\acp = 
{
{\Gamma(\overline{B} \to \overline{K}^*\gamma) - \Gamma(\bkg)} \over
{\Gamma(\overline{B} \to \overline{K}^*\gamma) + \Gamma(\bkg)}
},
\label{eq:acp}
\end{equation}

\begin{equation}
\Delta_{0-} =
{
{\Gamma(\overline{B}^0 \to \overline{K}^{*0}\gamma) - \Gamma(B^{-}\to K^{*-}\gamma)} \over
{\Gamma(\overline{B}^0 \to \overline{K}^{*0}\gamma) + \Gamma(B^{-}\to K^{*-}\gamma)}
}.
\label{eq:isospin}
\end{equation}

\noindent
The $\kszksz$ mode is excluded from the determination of the $CP$ asymmetry.  Being more precise, these quantities allow the SM to be more stringently tested.  The SM predictions for the $CP$ asymmetry~\cite{Greub:1994} are on the order of 1\%, while the isospin asymmetry~\cite{Matsumori:2005ds, Kagan:2001zk} ranges from 2-10\%.  The experimental measurements (Table~\ref{tab:prevMeasurements}) are in good agreement with these predictions.  However, new physics models could alter the SM estimates significantly~\cite{Kagan:2001zk, Ahmady:2006gh, Dariescu:2007}, and thus precise measurements constrain new physics parameter space. 

This note reports on a measurement of the branching fractions ${\cal B}(\bkgneut)$ and ${\cal B}(\bkpg)$, the isospin asymmetry $\isoas$, and the direct $CP$ asymmetries, $\acp(\bkgneut)$ and $\acp(\bkpg)$.

\begin{table}[h]
\begin{center}
   \begin{tabular}{l|r|r|r}
    \hline\hline
                                       & CLEOII \cite{Coan:1999kg}     & $\babar$ \cite{Dasu:2004kg} & \textsl{Belle}\cite{Nakao:2004kg}  \\
                                   & $9.2fb^{-1}$                  & $81.9fb^{-1}$                 & $78.0fb^{-1}$                    \\ \hline
     $\bkgneut$  & $4.55^{+0.72}_{-0.68}\pm0.34$ & $3.92\pm0.20\pm0.24$          & $4.01\pm0.21\pm0.17$             \\ 
     $(\times 10^{-5})$            &                               &                               & \\ \hline 
     $\bkpg$                       & $3.76^{+0.89}_{-0.83}\pm0.28$ & $3.87\pm0.28\pm0.26$          & $4.25\pm0.31\pm0.24$             \\ 
     $(\times 10^{-5})$            &                               &                               & \\ \hline 
     $\acp$                        & $+0.08\pm0.13\pm0.03$         & $-0.013\pm0.036\pm0.010$      & $-0.015\pm0.044\pm0.012$         \\ \hline
     $\Delta_{0-}$                 & N/A                           & $+0.050\pm0.045\pm0.028\pm0.024$ & $+0.012\pm0.044\pm0.026$   \\ \hline\hline
    
   \end{tabular}
   \caption{Previous measurements of the branching ratios and asymmetries.  The first and second errors are statistical and systematic respectively.  The last error on the isospin asymmetry for the $\babar$ measurement refers to the error on the production ratio of charged to neutral B events, $R^{+/0} \equiv \Gamma(\Upsilon(4S)\rightarrow B^+B^-)/\Gamma(\Upsilon(4S)\rightarrow B^0\bar{B^0})$.}
   \label{tab:prevMeasurements}  
\end{center}
\end{table}

\section{THE \babar\ DETECTOR AND DATASET}
\label{sec:babar}

We use a data sample containing \numBB~\BB\ events, corresponding to an integrated luminosity of $\lumitotal$ collected at the $\Upsilon(4S)$ resonance, taken with the \babar\ detector at the \pep2\ asymmetric-energy \epem\ collider located at the Stanford Linear Accelerator Center (SLAC).  These results supersede the previous \babar\ measurements~\cite{Dasu:2004kg}.

The \babar\ detector is described in Ref.~\cite{BabarDet}.  
Two components that are especially important for this analysis are the CsI 
Electromagnetic Calorimeter (EMC), used to identify and measure
photon energies, and the DIRC Cherenkov detector, used to identify charged
particles.

\section{ANALYSIS METHOD}
\label{sec:Analysis}

We reconstruct $\bkgneut$ using the modes $\kszkp$ and $\kszksz$, and $\bkpg$ using the decay modes $\kspkpz$ and $\kspksp$.  A high energy photon is combined with each vector meson.

The dominant source of background is continuum events ($\ep\en \to q\bar{q}$, with $q=u,d,s,c$) that contain a
high-energy photon from $\piz$ or $\eta$ decay.  The remaining background consists primarily of initial-state radiation (ISR) processes, and higher-multiplicity $\b\rightarrow\s\gamma$ decays, where one or more particles has not been reconstructed.  In addition, the decays of $\bkg$ can enter the signal selection by mis-reconstructing a similar mode.  For example, the decay $\bkg$($\kspkpz$) provides background for the mode $\bkg$($\kszkp$) by not correctly reconstructing the $\pi$ meson.  For each signal decay mode, selection requirements described below have been optimized for maximum statistical sensitivity with an assumed signal branching fraction of $4.0\times 10^{-5}$~\cite{Dasu:2004kg}.

Photon candidates are identified as localized energy deposits in the EMC that are not associated with any charged track.
The primary photon candidate is required to have a center-of-mass (CM) energy between 
1.5 and 3.5 \gev, to be well-isolated and have a shower shape
consistent with an individual photon~\cite{BabarksgOld}.  
In order to veto photons from $\piz$ and $\eta$ decays,
we form photon pairs composed of the signal photon candidate and all 
other photon candidates in the event.  
We then reject primary photon candidates consistent 
with coming from a $\piz$ or $\eta$ decay 
based on a likelihood ratio that uses the energy of
the partner photon, and the invariant mass of the pair.

The charged tracks must be well-reconstructed in the 
drift chamber, and are required to 
be consistent with coming from the \ep\en interaction region.  
They are identified as $K$ or $\pi$ mesons by the Cherenkov angle with 
respect to track direction, as well as by energy loss of the track (dE/dx).  
The $K_S$ candidates are reconstructed from two oppositely charged tracks 
that come from a common vertex.  
We require the invariant mass of the pair to be 
$0.49 < m_{\pi^{+}\pi^{-}} < 0.52\GeVcc$($0.48 < m_{\pi^{+}\pi^{-}} < 0.52\GeVcc$) 
and have a $K_S$ flight length significance requirement of 9.3(10) for the 
$\kszksz$($\kspksp$) mode.  

We form $\piz$ candidates by combining two photons (excluding the primary photon candidate) in the event, each of which has an energy greater than 30 \mev in the laboratory frame.  
We require the invariant mass of the pair to be 
$0.112 < m_{\gamma\gamma} < 0.15\GeVcc$ and 
$0.112 < m_{\gamma\gamma} < 0.15\GeVcc$ for the $\kszksz$ and $\kspkpz$ modes respectively.    
In order to refine the $\piz$ three momentum vector, we perform a mass-constrained fit 
of the two photons.  

We combine the reconstructed $K$ or $\pi$ mesons to form $K^*$ candidates.  
We require the invariant mass of the pair to satisfy 
$0.78 < m_{K^{+}\pi^{-}} < 1.1\GeVcc$, 
$0.82 < m_{K_{S}\pi^{0}} < 1.0\GeVcc$, 
$0.79 < m_{K^{+}\pi^{0}} < 1.0\GeVcc$, 
and $0.79 < m_{K_{S}\pi^{+}} < 1.0\GeVcc$. 
The charged track pairs are required to originate 
from a common vertex consistent with the \epem collision region.   

We combine the $K^*$ and high-energy photon candidates to form $B$ candidates. 
We define in the CM frame (the asterisk denotes the CM quantity) 
$\de \equiv E^*_{B}-E_{\rm beam}^*$, where $E^*_B$ is the energy of the 
$B$ meson candidate and $E_{\rm beam}^*$ is the beam energy.  
We also define the beam-energy-substituted mass 
$\mes \equiv \sqrt{ E^{*2}_{\rm beam}-{\mathrm{{p}}}_{B}^{\;*2}}$, 
where ${\mathrm{{p}}}_B^{\;*}$ is the momentum of the $B$ candidate.  
In addition, we consider the helicity angle $\theta_{H}$ of the $K^*$, defined as the angle between one of the daughters of the 
$K^*$ meson and the $B$ candidate in the $K^*$ rest frame.     
Signal events have $\de$ close to zero 
with a resolution of approximately $50\mev$, and an $\mes$ distribution 
centered at the mass of the $B$ meson with a resolution of $3~\mevcc$.  
Since the $K^*$ recoils against a photon, it has a 
$\cos\theta_{H}$ distribution of $\sin^2\theta$.  
We only consider candidates in the ranges $-0.3 < \de <0.3 \GeV$, 
$\mes >  5.22\GeVcc$, and $|\cos\theta_{H}|< 0.75$.  
The latter selection is to reject background such as $B \to K^{*}\eta$ 
and $B \to K^{*}\piz$, which are distributed as $\cos^2\theta$ 
in $\cos\theta_{H}$.
To ensure the events are properly reconstructed, we apply
a selection criterion to the separation (and its uncertainty) along the beam axis 
between the B meson candidate and the rest of the event (ROE).  The ROE is defined as all charged tracks and neutral energy deposits in 
the calorimeter that are not used to reconstruct the B candidate.

In order to reject continuum background, we combine 13
variables into a neural network (NN)~\cite{Haykin:1994}.  
One class of these variables exploits the topological differences 
between spherical signal events and jet-like continuum events by 
considering information from the B meson candidate and the ROE.
The other class exploits the difference in particle 
production mechanisms between B meson decays and continuum events.     
The discriminating variables are described in 
Ref.~\cite{rhogamma}.  
Each mode has a separately trained neural network.  
The output of this network peaks at a value of one for signal-like events. 
We select events with a criterion on this output that is optimized for
maximum statistical sensitivity. 
To validate the neural network, we use a $\bdpi$ control sample.

After applying all the selection criteria, we select the best candidate in each event 
by choosing the candidate with the reconstructed $K^*$ mass closest 
to the nominal mass.  
On average, across all four modes, there are 
approximately 1.1 candidates per event in signal events.

We perform an unbinned maximum likelihood fit to extract 
the signal yield, constructing a separate fit for each mode.  
We use three observables ($\mes$, $\de$, and $\cos\theta_{H}$) for each 
candidate event and assume three hypotheses (signal, continuum, and $\BB$)
from which the candidate can originate.  All $\BB$ background is included in the $\BB$ component. 
The use of $\cos\theta_{H}$ suppresses 
the $\BB$ background.  Since the correlations among the three dimensions are small, we use uncorrelated probability distribution functions (PDF) to construct the likelihood function.  The correction to this method is determined in section~\ref{sec:Systematics}.  The likelihood function is:

$$
{\cal L}=\exp{\left(-\sum_{i = 1}^{M} n_{i}\right)}\cdot
\left(\prod_{j = 1}^{N} \left[\sum_{i=1}^M n_i{\cal P}_i(\vec{x}_j;\
\vec{\alpha}_i)\right]\right)
$$

\noindent
where N is the number of events, M is the number of hypotheses, 
$n_{i}$ represents the yield of a particular hypothesis, 
${\cal P}_i(\vec{x}_j;\vec{\alpha}_i)$  is the product of 
one-dimensional PDFs over 
the three dimensions, 
$\vec{x}_j = (\mes,\de,\cos\theta_{H})$, 
and the $\vec{\alpha}_i$ represent the fit parameters.

The signal $\mes$ PDF for the $\kszkp$ and $\kspksp$ modes is parameterized as 

\begin{equation}
f(x) = \exp \left[ \frac{-(x-\mu)^2}{2 \sigma^2_{L,R} + \alpha_{L,R} (x-\mu)^2} \right], 
\label{eq:cruijff}
\end{equation}

\noindent
where $\mu$ is the peak position of the distribution, 
$\sigma_{L,R}$ are the widths to the left and right of the peak, 
and $\alpha_{L,R}$ are a measure of the tails to the left 
and right of the peak, respectively.  
We constrain $\sigma_{L} = \sigma_{R}$, and fix $\alpha_{L,R}$ to the 
values obtained from Monte Carlo (MC) simulation~\cite{geant}.  
For the $\kszksz$ and $\kspkpz$ modes, the signal $\mes$ distribution is 
described by a Crystal Ball function~\cite{CrysBall}.  
The Crystal Ball function has a single tail parameter, $\alpha$, 
which we fix to the value determined from MC.
For each mode, the signal $\de$ distribution is described by the same 
function in Eq.~\ref{eq:cruijff}, but with different values for the parameters.  
However, we allow $\sigma_{L}$ and $\sigma_{R}$ to float independently, 
but still fix the values of $\alpha_{L,R}$ to MC.  
For all components, the $\cos\theta_{H}$ distribution is 
modeled by a low order polynomial, which is fixed to the MC values.  
For the continuum hypothesis, the $\mes$ PDF is parameterized 
by an ARGUS function~\cite{argus}, with its shape parameter floating in the fit.  
The continuum $\de$ shape is modeled by a low order polynomial with its parameters
floating in the fit. 
Various functional forms are used to describe the $\BB$ background,
all parameters of which are taken from MC simulation and held fixed.

The CP asymmetry $\acp$ parameter is measured in the three  ``self-tagging'' modes:
$\kszkp$, $\kspkpz$ and $\kspksp$. 
The fit is accomplished by performing a simultaneous fit to the two flavor 
sub-samples ($K^*$ and $\overline{K^*}$) in each mode.
All shape parameters are assumed to be flavor independent and the $\acp$
of each component is floated in the fit.

\begin{figure}[]
 \begin{center}
  \resizebox{0.48\textwidth}{!}{\includegraphics{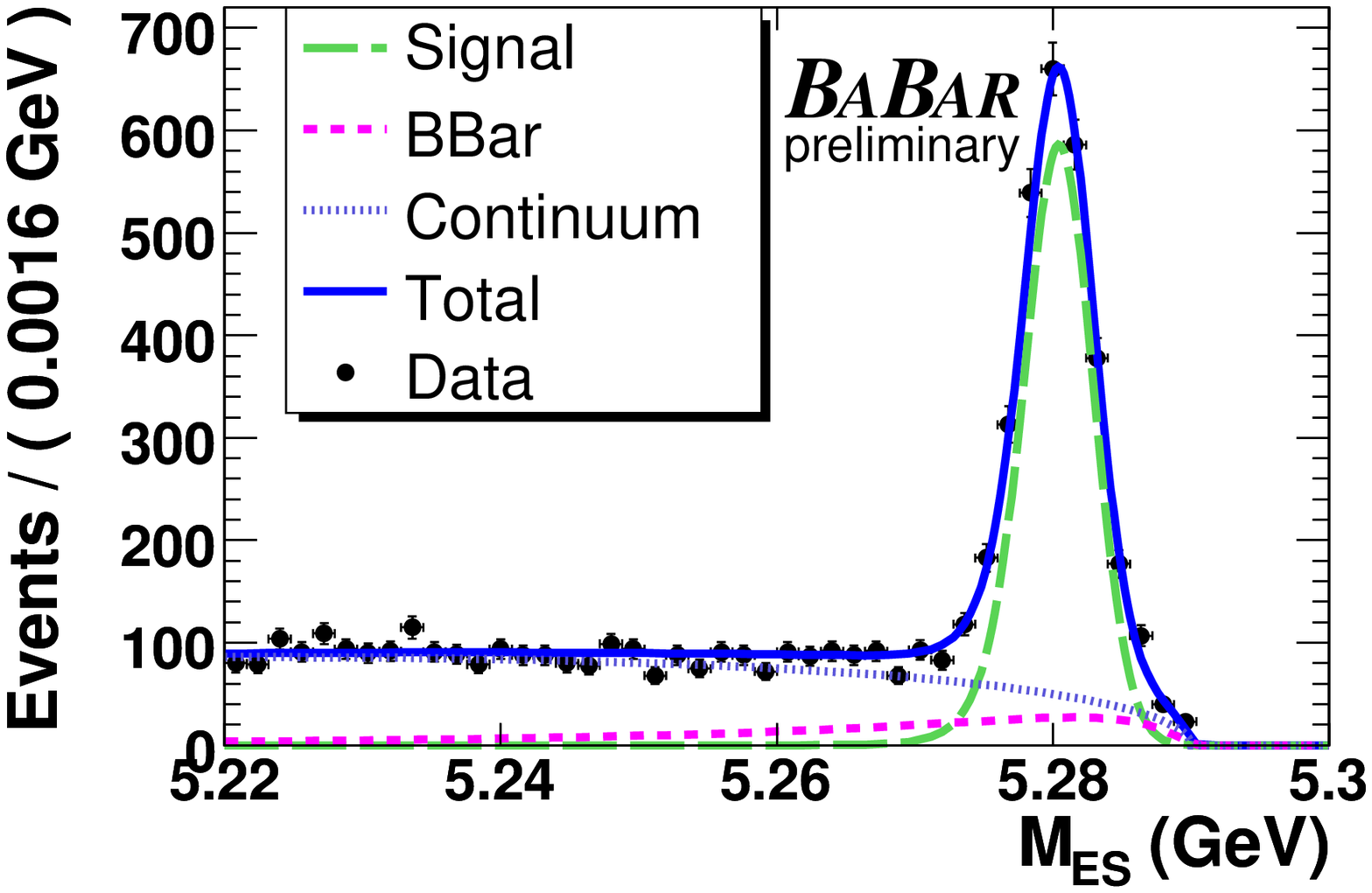}}
  \resizebox{0.48\textwidth}{!}{\includegraphics{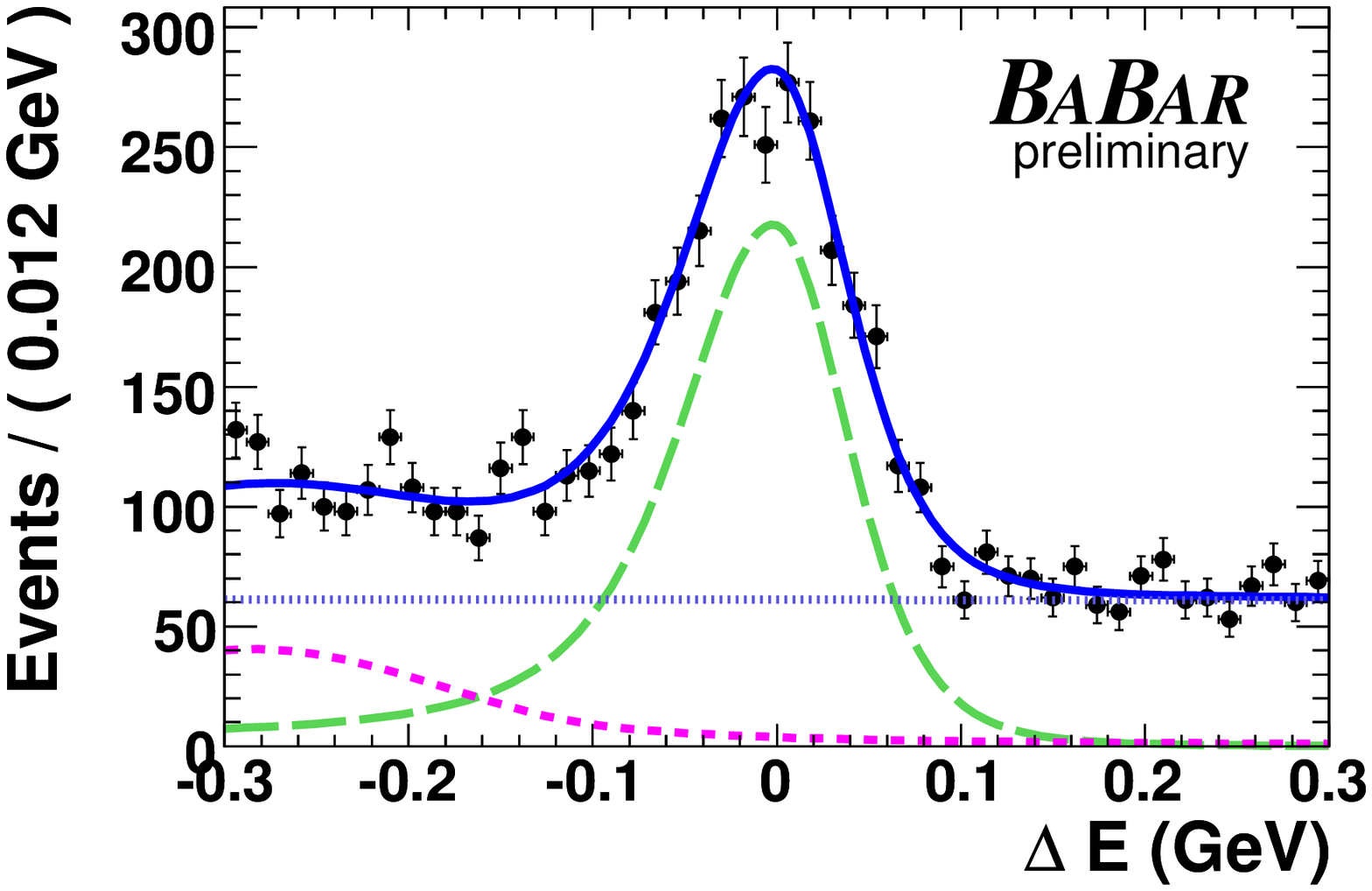}}
  \resizebox{0.48\textwidth}{!}{\includegraphics{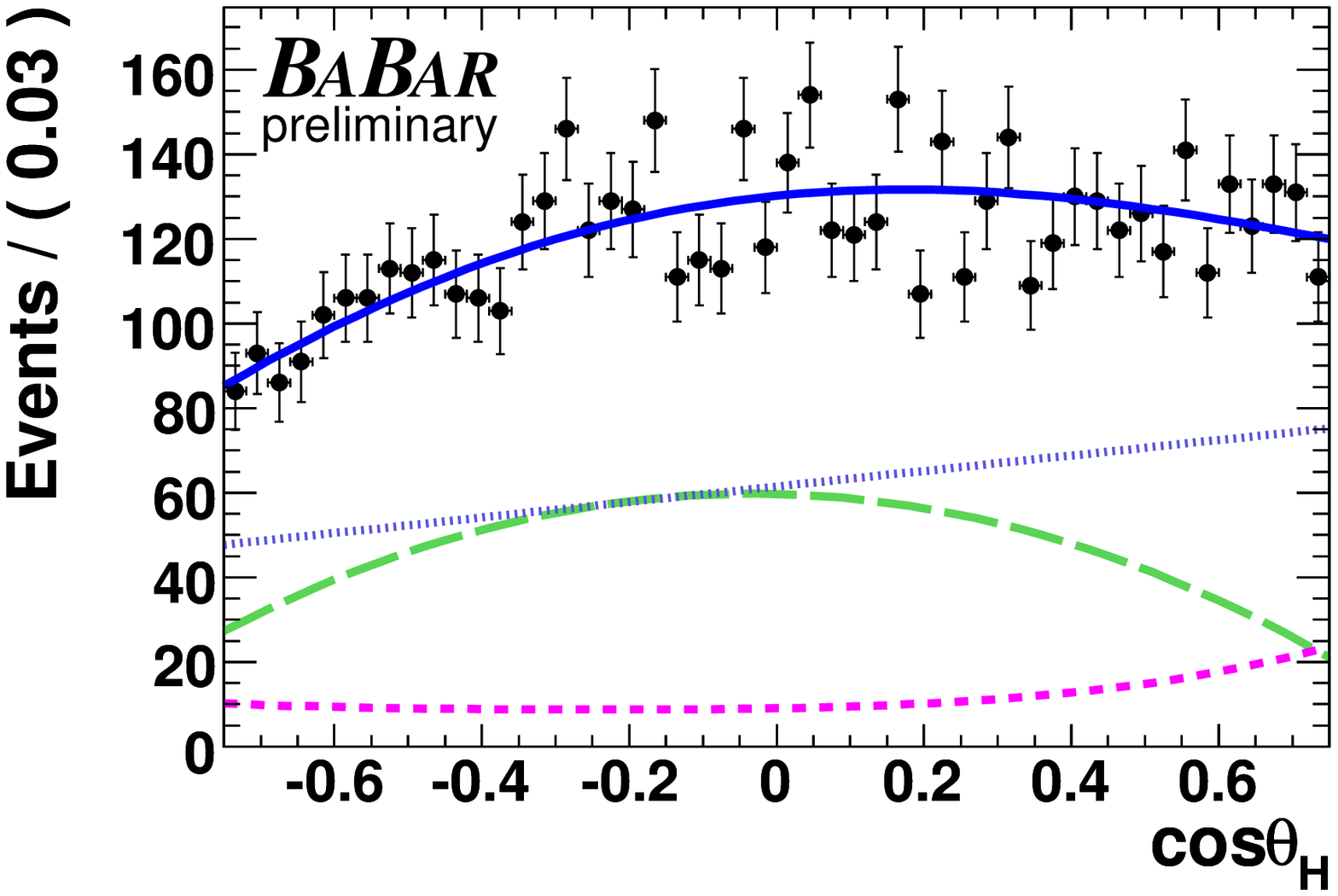}}
  \caption[$\kszkp$ projection plots of the full fit to data]
    {$\kszkp$ projection plots of the full fit to data.  The daughter of the $K^*$ used to determine the helicity angle is the $K$ meson. }
  \label{fig:fullfit_kpi_data}
 \end{center}
\end{figure}

\begin{figure}[]
 \begin{center}
  \resizebox{0.48\textwidth}{!}{\includegraphics{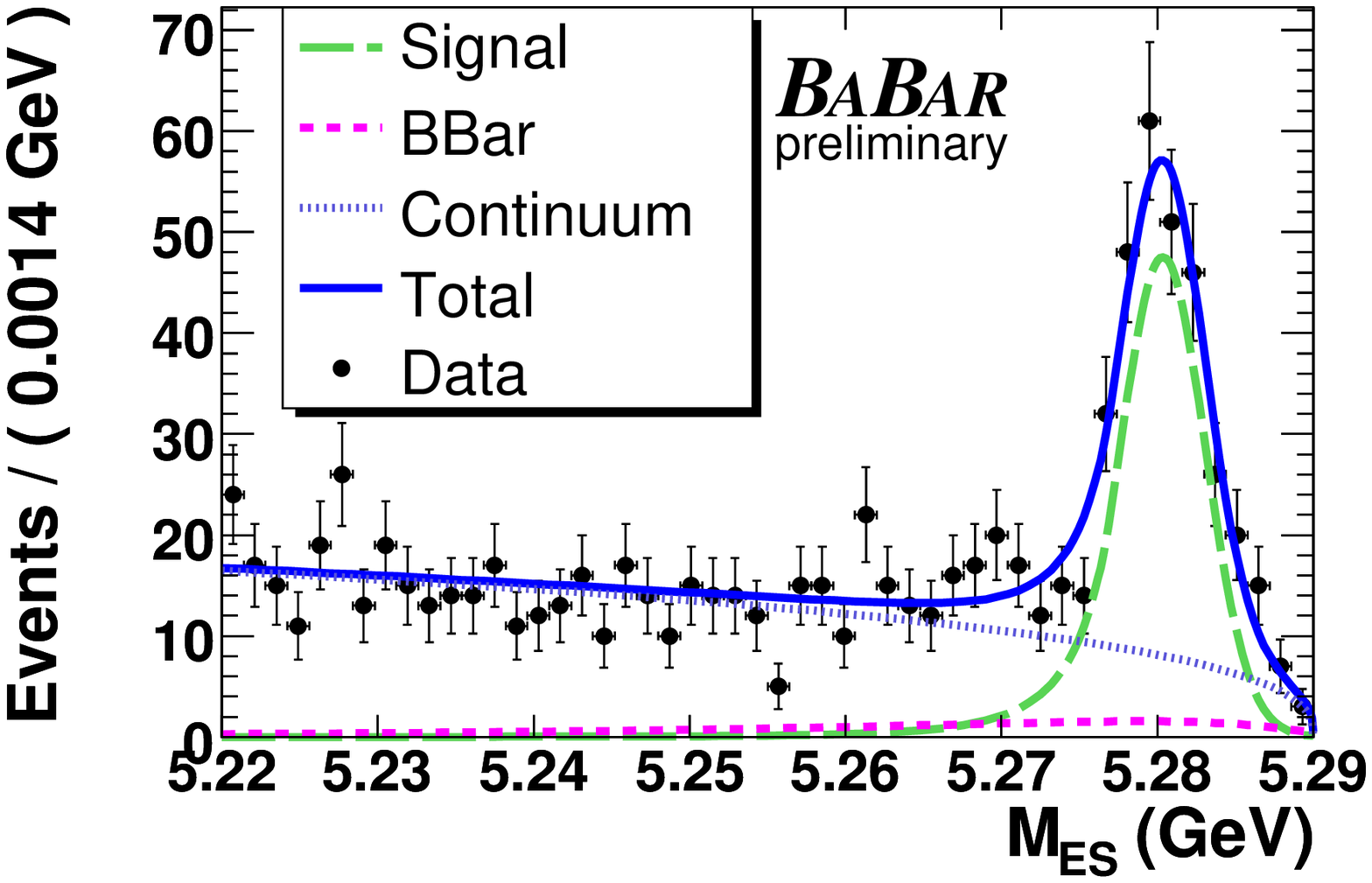}}
  \resizebox{0.48\textwidth}{!}{\includegraphics{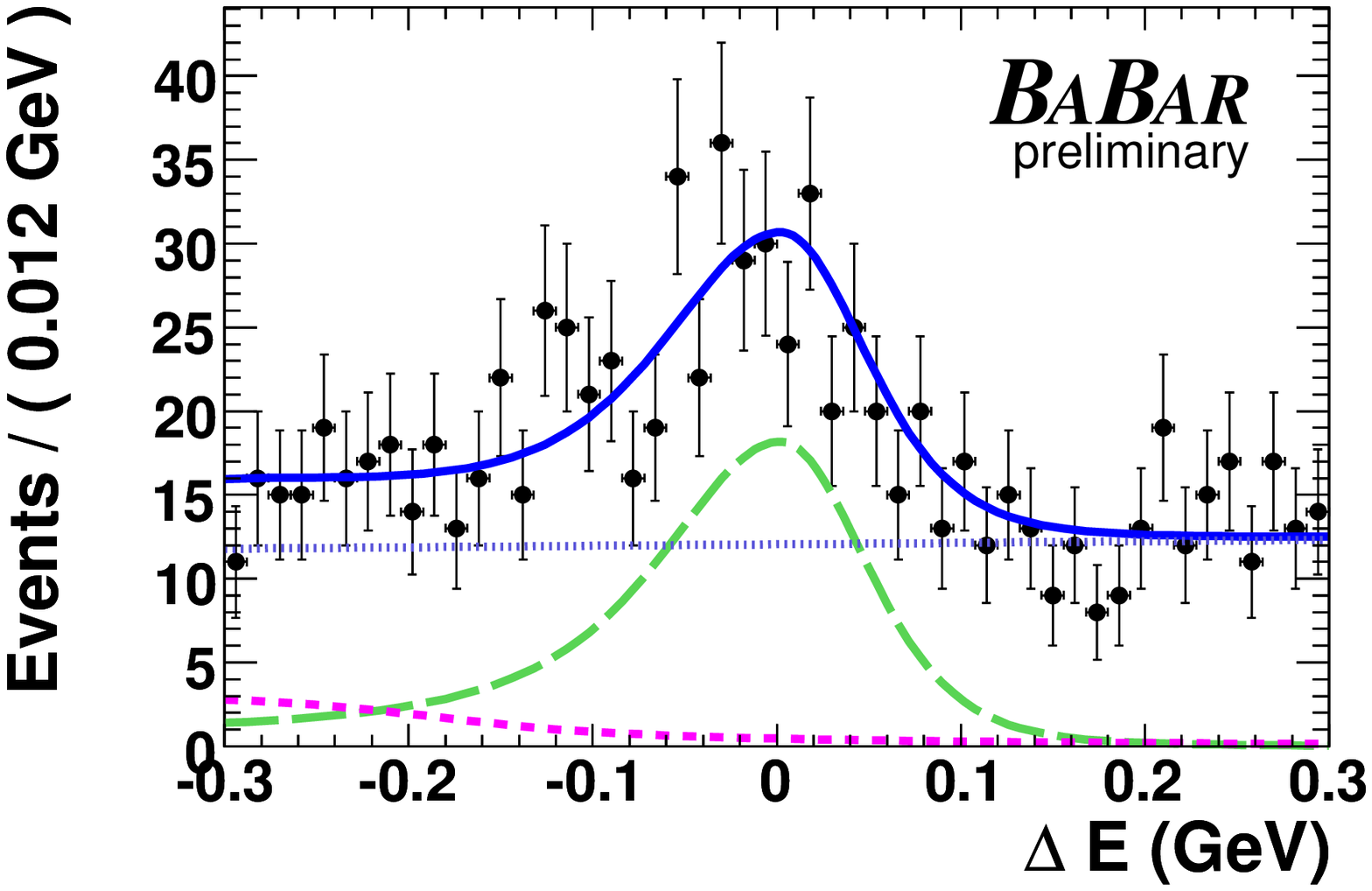}}
  \resizebox{0.48\textwidth}{!}{\includegraphics{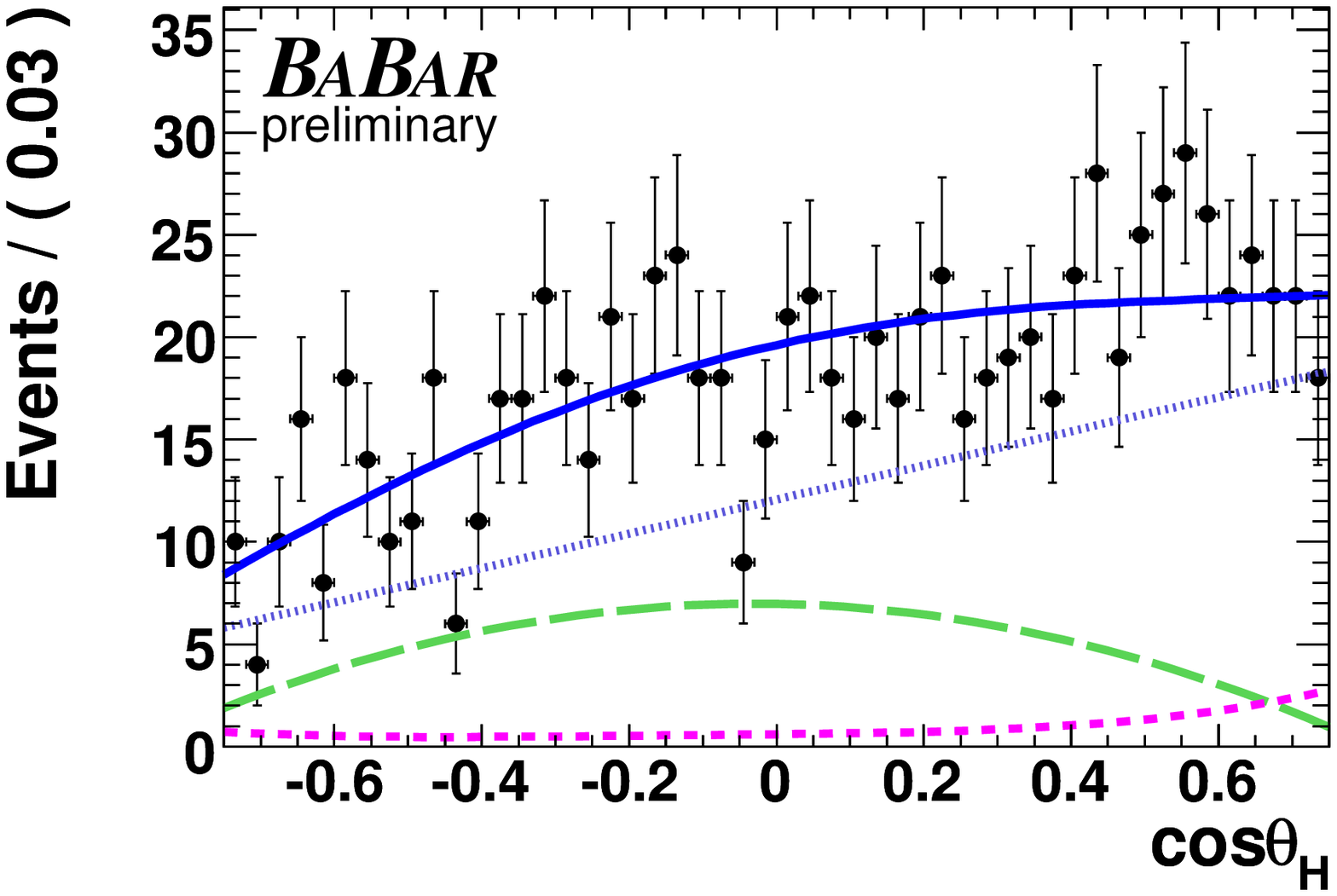}}
  \caption[$\kszksz$ projection plots of the full fit to data]
    {$\kszksz$ projection plots of the full fit to data.  The daughter of the $K^*$ used to determine the helicity angle is the $K_S$. }
  \label{fig:fullfit_kspi0_data}
 \end{center}
\end{figure}

\begin{figure}[]
 \begin{center}
  \resizebox{0.48\textwidth}{!}{\includegraphics{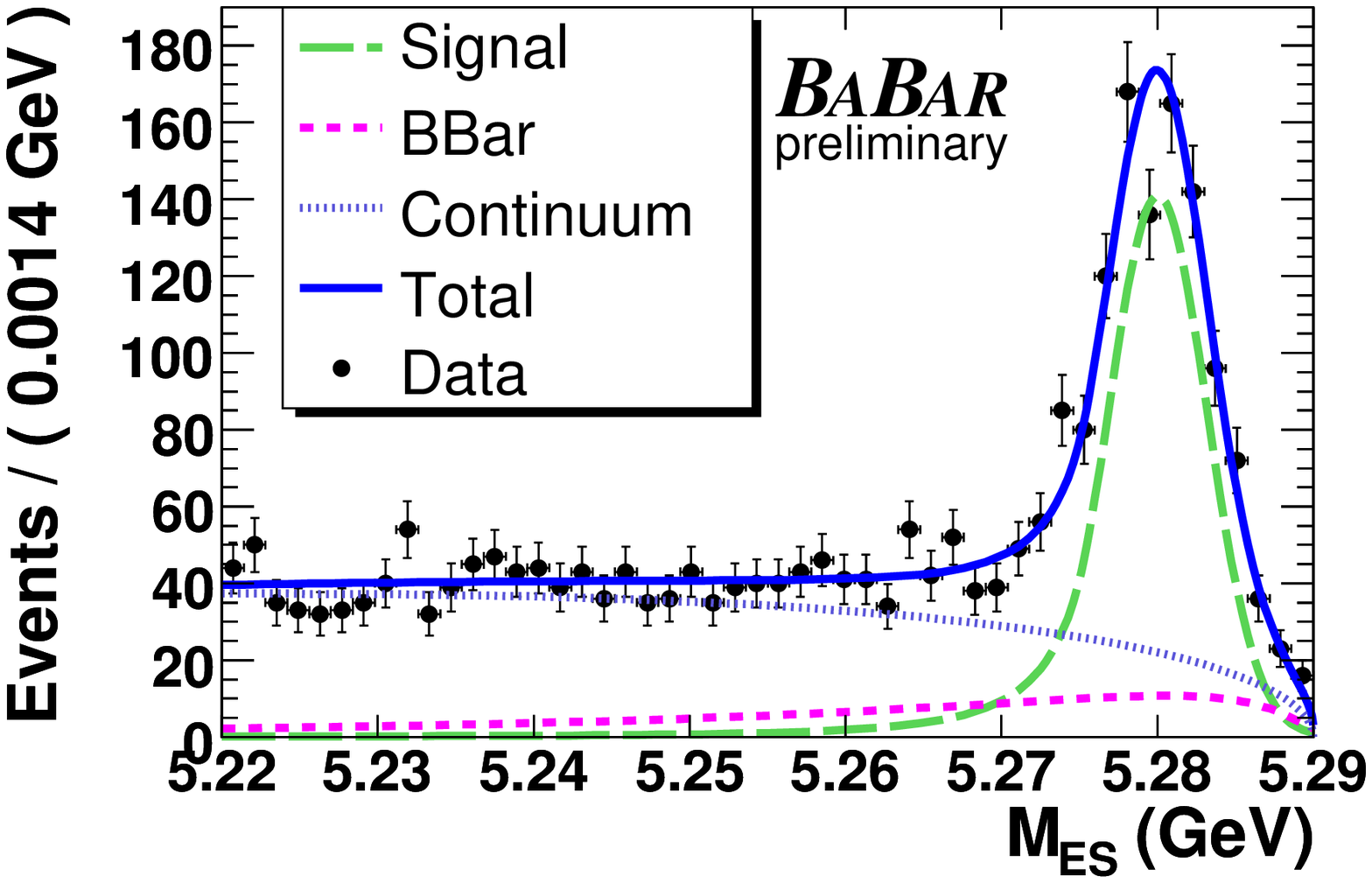}}
  \resizebox{0.48\textwidth}{!}{\includegraphics{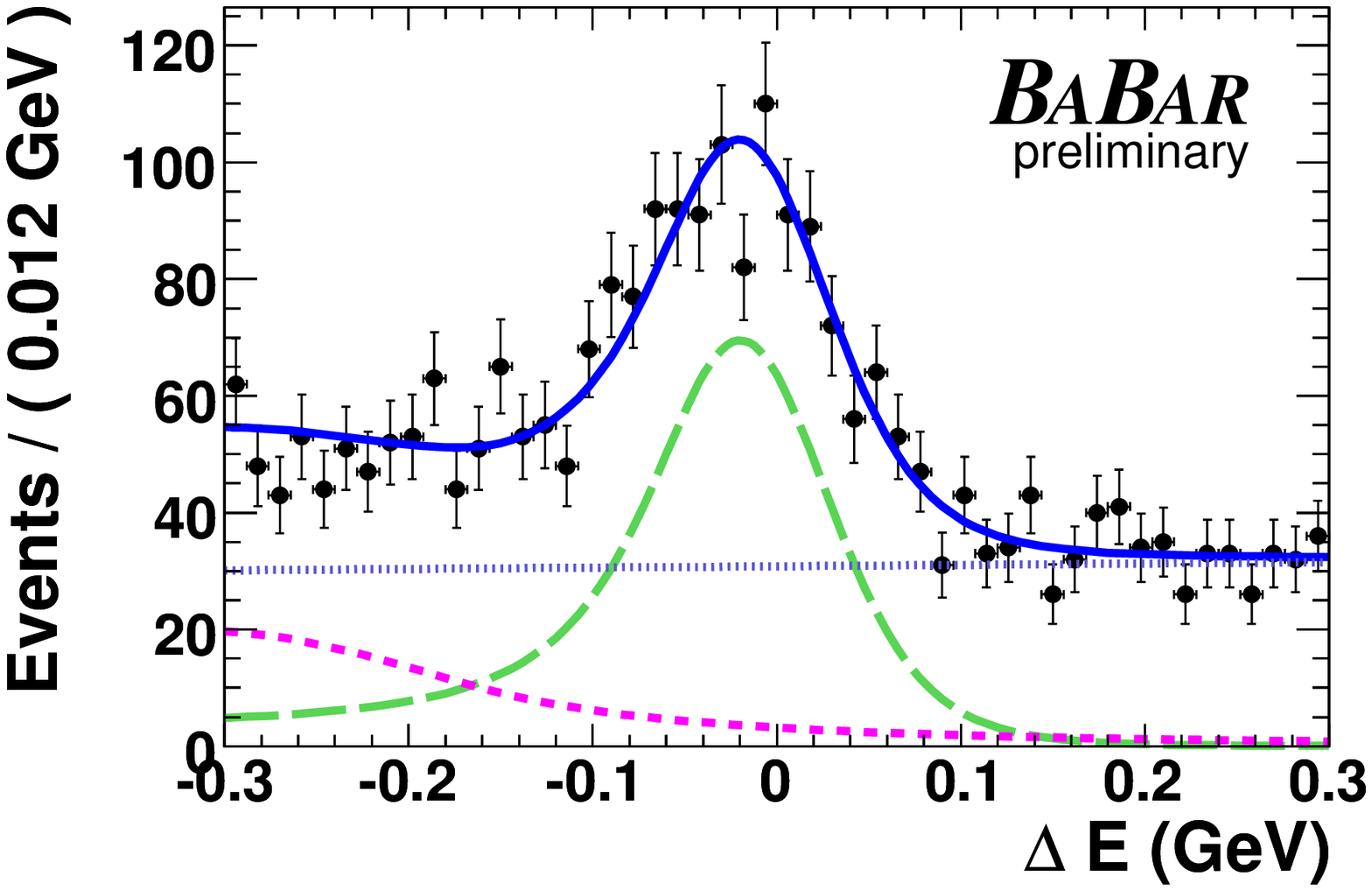}}
  \resizebox{0.48\textwidth}{!}{\includegraphics{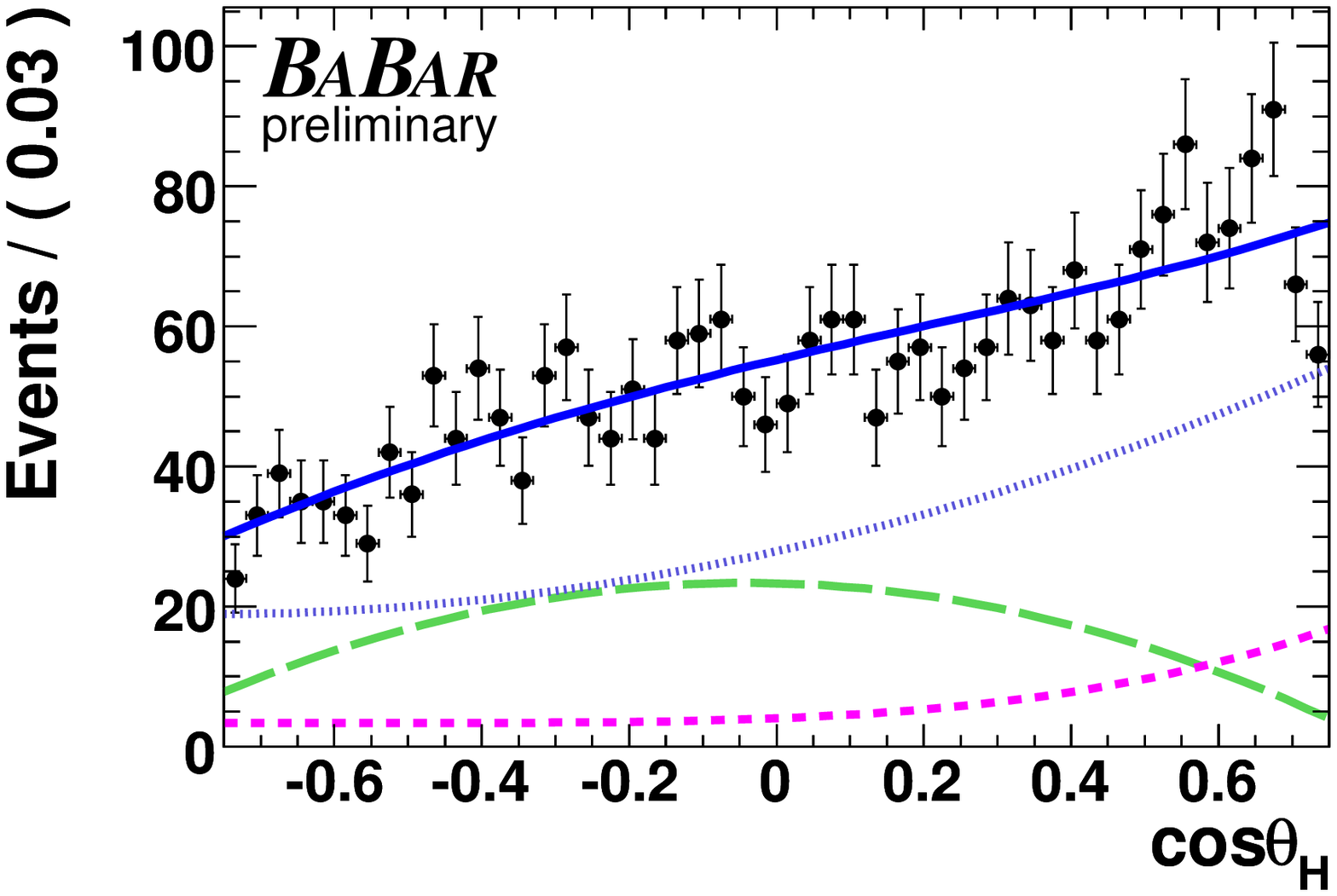}}
  \caption[$\kspkpz$ projection plots of the full fit to data]
    {$\kspkpz$ projection plots of the full fit to data.  The daughter of the $K^*$ used to determine the helicity angle is the $K^+$.   }
  \label{fig:fullfit_kpi0_data}
 \end{center}
\end{figure}

\begin{figure}[]
 \begin{center}
  \resizebox{0.48\textwidth}{!}{\includegraphics{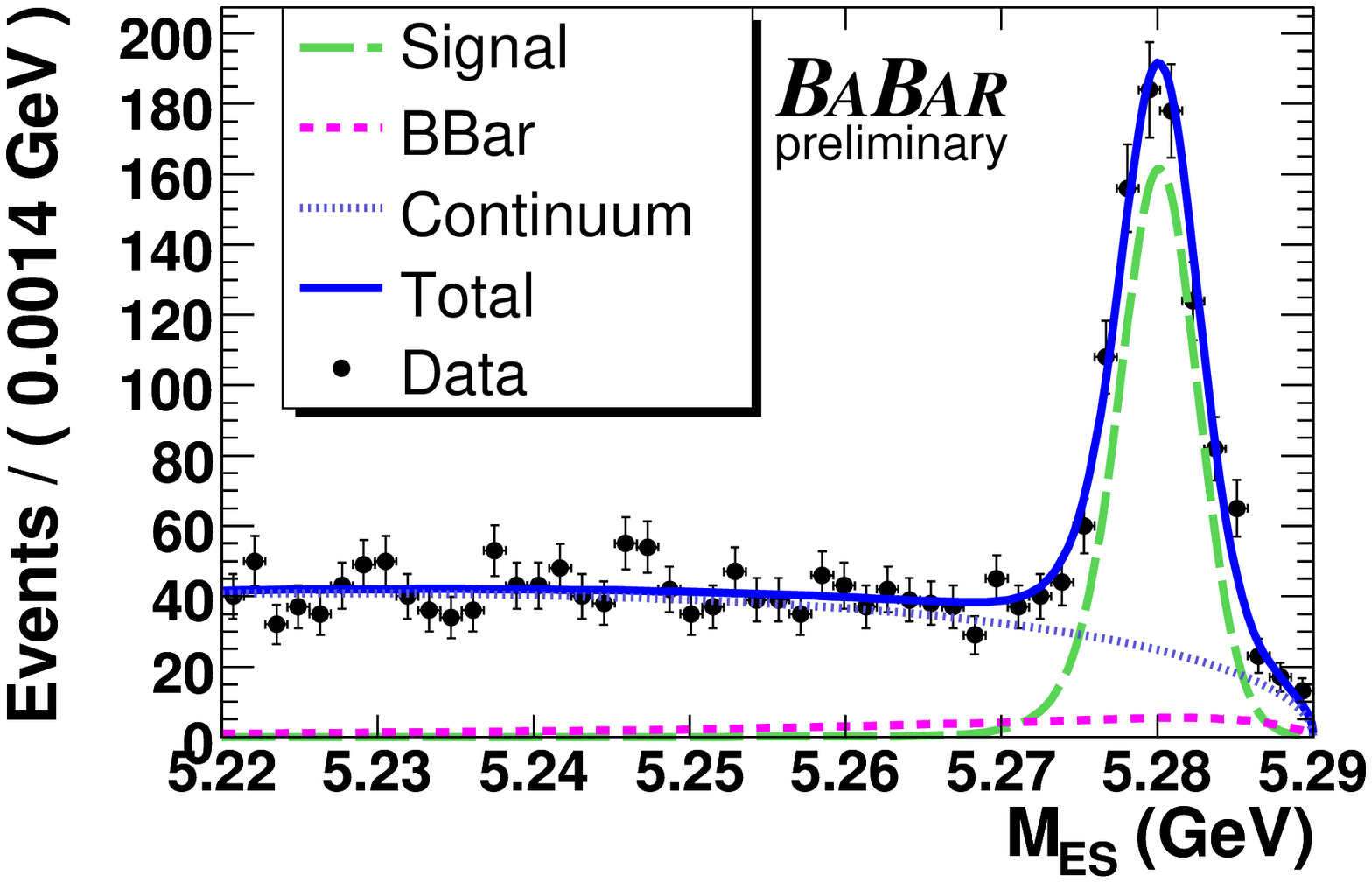}}
  \resizebox{0.48\textwidth}{!}{\includegraphics{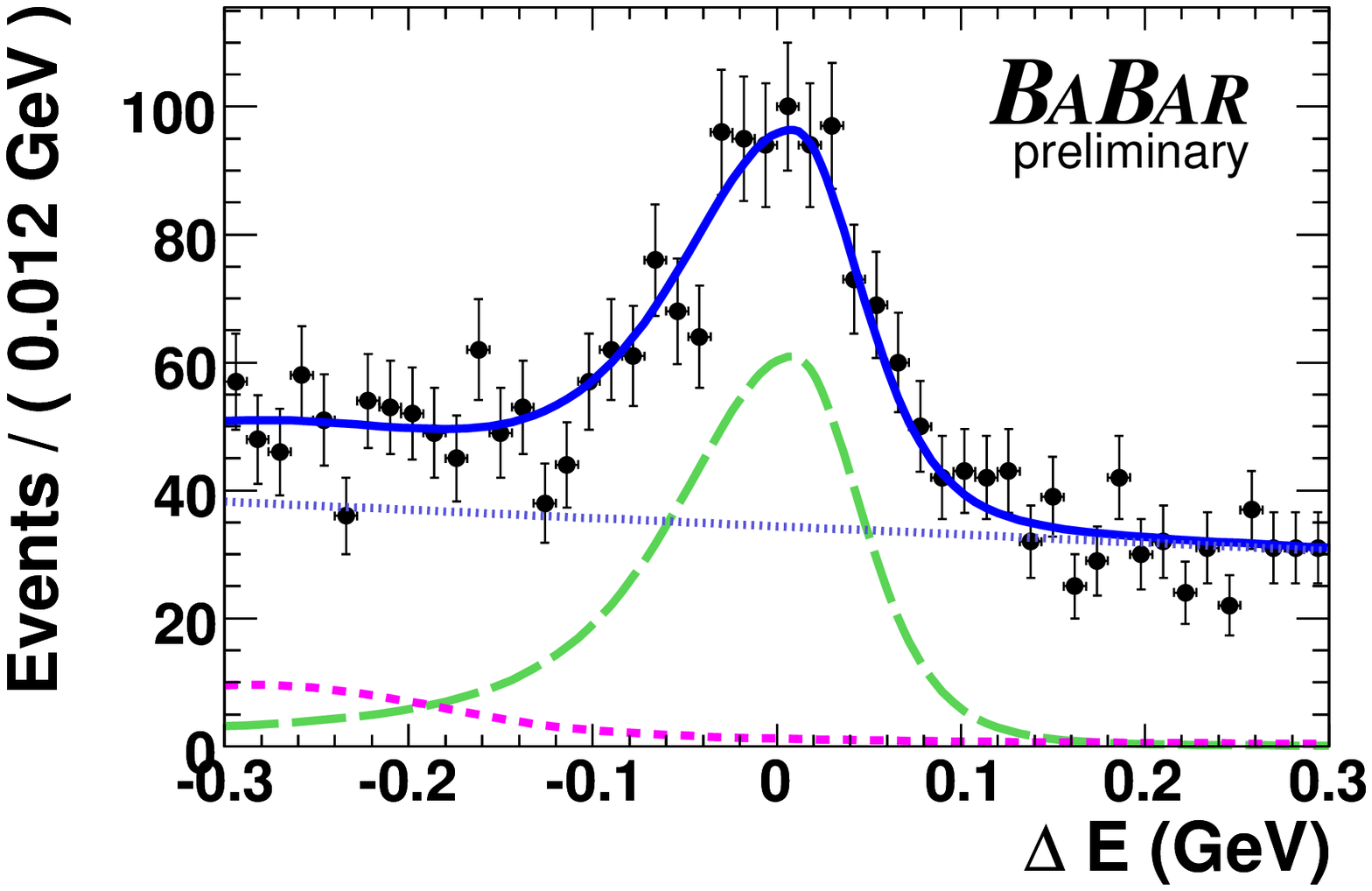}}
  \resizebox{0.48\textwidth}{!}{\includegraphics{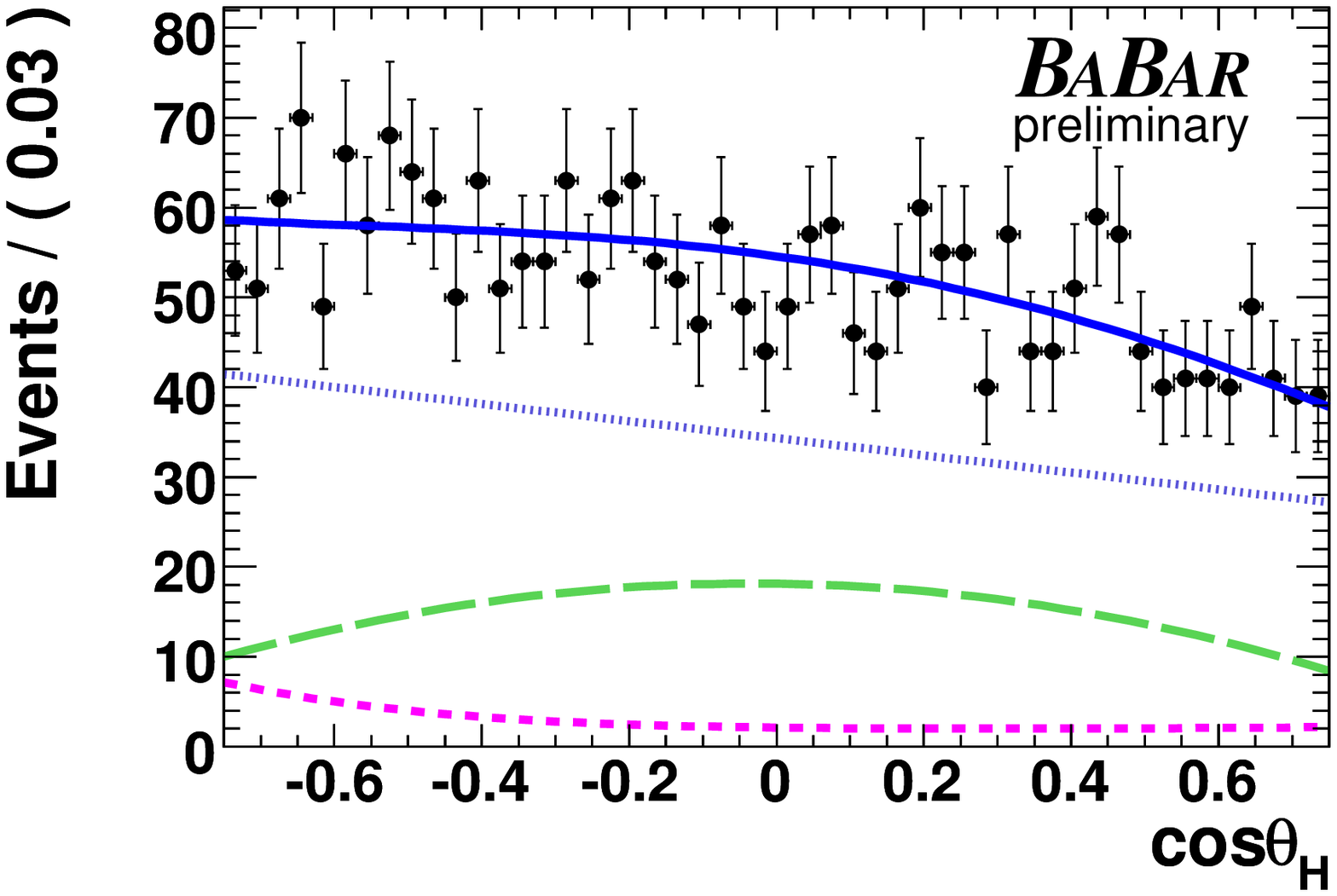}}
  \caption[$\kspksp$ projection plots of the full fit to data]
    {$\kspksp$ projection plots of the full fit to data.  The daughter of the $K^*$ used to determine the helicity angle is the $\pi$ meson.}
  \label{fig:fullfit_kspi_data}
 \end{center}
\end{figure}

Figures~\ref{fig:fullfit_kpi_data} through~\ref{fig:fullfit_kspi_data} 
show the projections of the likelihood fit to data.   For each projection, signal region cuts ($5.27 < \mes < 5.29\GeVcc$, $-0.2 < \de < 0.1 \GeV$) have been applied, except the $\mes$ selection is not applied to the $\mes$ distribution and similarly for $\de$.  The asymmetry of the signal component of the $\cos\theta_{H}$ distributions is due to mis-reconstructed signal candidates.
Table~\ref{table:bfresults} shows the results for the branching fractions 
and $CP$ asymmetry, where the sign of $\acp$ is defined by Eq.~\ref{eq:acp}.

\begin{table}[htbp]

        \begin{center}

        \begin{tabular*}{\linewidth}{
@{\extracolsep{\fill}}l
@{\extracolsep{\fill}}c
@{\extracolsep{\fill}}c
@{\extracolsep{\fill}}c
@{\extracolsep{\fill}}c
}\hline\hline

    Mode&
    {$\epsilon$(\%)}&
    {$N_S$}&
    {${\cal B}(\times\mbox{10}^{-\mbox{5}}) $}&
    {${\acp}$}     \\\hline

    $K^+\pi^-$&
    20.6$\pm$0.7&
    2394.1$\pm$55.6&
    $\kszkpbr \pm \kszkpbrstat \pm \kszkpbrsyst$&
    $-0.023 \pm 0.022 \pm 0.011$\\

    $K_s\pi^0$&
    11.7$\pm$0.8&
    \ 256.0 $\pm$20.6&
    $\kszkszbr \pm \kszkszbrstat \pm \kszkszbrsyst$&
    N/A\\

    $K^+\pi^0$&
    13.7$\pm$0.7&
    \ 872.7$\pm$37.6&
    $\kspkpzbr \pm \kspkpzbrstat \pm \kspkpzbrsyst$&
    $+0.033 \pm 0.039 \pm 0.011$\\

    $K_s\pi^+$&
    18.8$\pm$0.7&
    \ 759.1$\pm$33.8&
    $\kspkspbr \pm \kspkspbrstat \pm \kspkspbrsyst$&
    $-0.006 \pm 0.041 \pm 0.011$\\\hline

    $\bkgneut$&&&
    $\kszcombr \pm \kszcomstat \pm \kszcomsyst$&\\

    $\bkpg$&&&
    $\kspcombr \pm \kspcomstat \pm \kspcomsyst$&\\

    $\bkg$&&&& 
    $\acpval \pm \acpstat \pm \acpsyst$ \\
     
    \hline\hline

    \end{tabular*}
    \end{center}

\caption{The signal reconstruction efficiency $\epsilon$, the fitted signal yield $N_S$,  branching fraction ${\cal B}$, and $CP$ asymmetry ($\acp$) for each decay mode.  The signal efficiencies have been corrected for differences between the selection efficiency in data and MC.  Errors are statistical and systematic, with the exception of $\epsilon$ and $N_S$, which have only systematic and statistical errors respectively.  Also shown are the combined branching fractions and $CP$ asymmetry.}

\label{table:bfresults}
\end{table}

\section{SYSTEMATIC ERROR STUDIES}
\label{sec:Systematics}

Table~\ref{table:sys} lists the sources of systematic uncertainty 
for all four modes.  
These are associated with the signal reconstruction efficiency, 
modeling of the $\BB$ background, and the choice of fixed parameters of the fit PDFs.  
The ``Photon selection'' systematic error is a combination
of the photon efficiency, the isolation criteria,  and the shower shape selection.
For the Neural Net and the $\pi^0$/$\eta$ veto, 
we use a $\bdpi$ control sample  
to determine the systematic error.  
The ``Fit Model'' systematic error is a combination of incorporating uncertainties 
due to our imperfect knowledge of the 
normalization and shape of the inclusive $\incbsg$ spectra, 
and the choice of fixed parameters.  We also perform a series of experiments in which we select signal events from MC simulation and combine them with events from background generated using PDFs from the fit.  The bias resulting from correlations among the three dimensions, or the PDFs incorrectly modeling the signal distribution can be determined using this procedure.  The ``Signal PDF bias'' systematic error results from these series of experiments.
Associated with all of the systematic uncertainties 
is a correction factor, which is a ratio between the estimated efficiency in data and the corresponding efficiency in MC.  The corrections are 0.953, 0.897, 0.919, and 0.936 for the $\kszkp$, $\kszksz$, $\kspkpz$, and $\kspksp$ modes respectively.  We use this factor to correct the MC reconstruction efficiency.         

The systematics of the $\acp$ measurement were studied in detail in
Reference \cite{BabarksgOld}.
They were found to be due to uncertainties in the hadronic cross section 
asymmetry and to reconstruction asymmetries.  
Here, we simply adopt the value 1.1\%, which is a conservative estimate due 
to reconstruction improvements.

\begin{table}[ht]

        \begin{center}

        \caption{Systematic errors (in \%) of the branching fractions.}
         \label{table:sys}
        \begin{tabular*}{\linewidth}{
@{\extracolsep{\fill}}l
@{\extracolsep{\fill}}c
@{\extracolsep{\fill}}c
@{\extracolsep{\fill}}c
@{\extracolsep{\fill}}c
}\hline\hline

Mode                            &    $\kszkp$ & $\kszksz$ & $\kspkpz$ & $\kspksp$ \\\hline 
$B\overline{B}$ sample size     &      1.1  &  1.1  &  1.1  &  1.1 \\ 
Tracking efficiency             &      1.2  &  -    &  0.6  &  0.8 \\ 
Particle identification         &      0.6  &  -    &  0.6  &  0.2 \\ 
Photon selection                &      2.2  &  2.2  &  2.2  &  2.2 \\ 
$\pi^0$ reconstruction          &      3.0  &  -    &  3.0  &   -  \\
$\pi^0$ and $\eta$ veto         &      1.0  &  1.0  &  1.0  &  1.0 \\
$K_{S}$ reconstruction          &       -   &  0.7  &   -   &  0.7 \\
Neural Net efficiency           &      1.5  &  1.0  &  1.0  &  1.0 \\ 
Fit Model                       &      0.7  &  5.3  &  2.9  &  1.6 \\ 
Signal PDF bias                 &      0.9  &  2.2  &  1.6  &  1.4 \\\hline
Sum in quadrature               &      3.5  &  7.1  &  5.3  &  3.7  \\
\hline \hline
\end{tabular*}

\end{center}

\end{table}

\section{RESULTS}
\label{sec:Results}
                
For the branching fraction calculation, we assume the production ratio, $R^{+/0}$, is unity.  $R^{+/0}$ is defined as

$$
R^{+/0} = \frac{\Gamma(\Upsilon(4s)\rightarrow B^+B^-)}
{\Gamma(\Upsilon(4s)\rightarrow B^0\bar{B^0})}.
$$

\noindent
The measured branching fractions are shown in 
Table~\ref{table:bfresults}.  
The combined branching fractions are calculated from the sub-modes
using the method of least squares, taking into account correlated systematic errors.  
  
To calculate the isospin asymmetry $\Delta_{0-}$, we combine the branching fractions, the ratio of the 
$B^+$ and $B^0$ lifetime $\tau_{+}/\tau_{0}$, and the 
production ratio $R^{+/0}$ according to 

\begin{equation}
\Delta_{0-} = \frac{1}{2}(IR^{+/0}\frac{\tau^+}{\tau^0} -1),
\label{eq:delta}
\end{equation}

\noindent
where $I$ is

$$
I  = 
\frac
{{\cal B}({B^0} \to {K^{*0}}\gamma)}
{{\cal B}(B^{*-}\to K^{*-}\gamma)},
$$

\noindent
to obtain the isospin asymmetry

$$\Delta_{0-} = \isoval \pm \isostat \pm \isosyst \pm \prodratio.$$

\noindent
The first and second errors are statistical and systematic, respectively.  The last error comes from the error on the production ratio, and we have used $\tau_{+}/\tau_{0} = 1.071 \pm 0.009$~\cite{PDG:2006}, $R^{+/0} = 1.020 \pm .034$~\cite{HFAG:2006}.  In addition, to obtain Eq.~\ref{eq:delta}, we have used the approximation that $I$, $R^{+/0}$, and $\tau_{+}/\tau_{0}$ are all close to unity.  The 90\% confidence interval for $\Delta_{0-}$ including systematic uncertainties is

 $$\isolow < \Delta_{0-} < \isohi.$$ 

\noindent
The corresponding time-integrated $CP$ asymmetry (table~\ref{table:bfresults}) is

  $$\acp = \acpval \pm \acpstat \pm \acpsyst,$$
  
\noindent
while the 90\% confidence interval for $\acp$ is

 $$\acplolim < \acp < \acphilim.$$ 

\noindent
The combined asymmetries are calculated using the same method as the branching fractions.

\begin{figure}[]
\noindent
 \begin{minipage}[t]{0.5\textwidth}
        \begin{center}
        \includegraphics[width = 80mm]{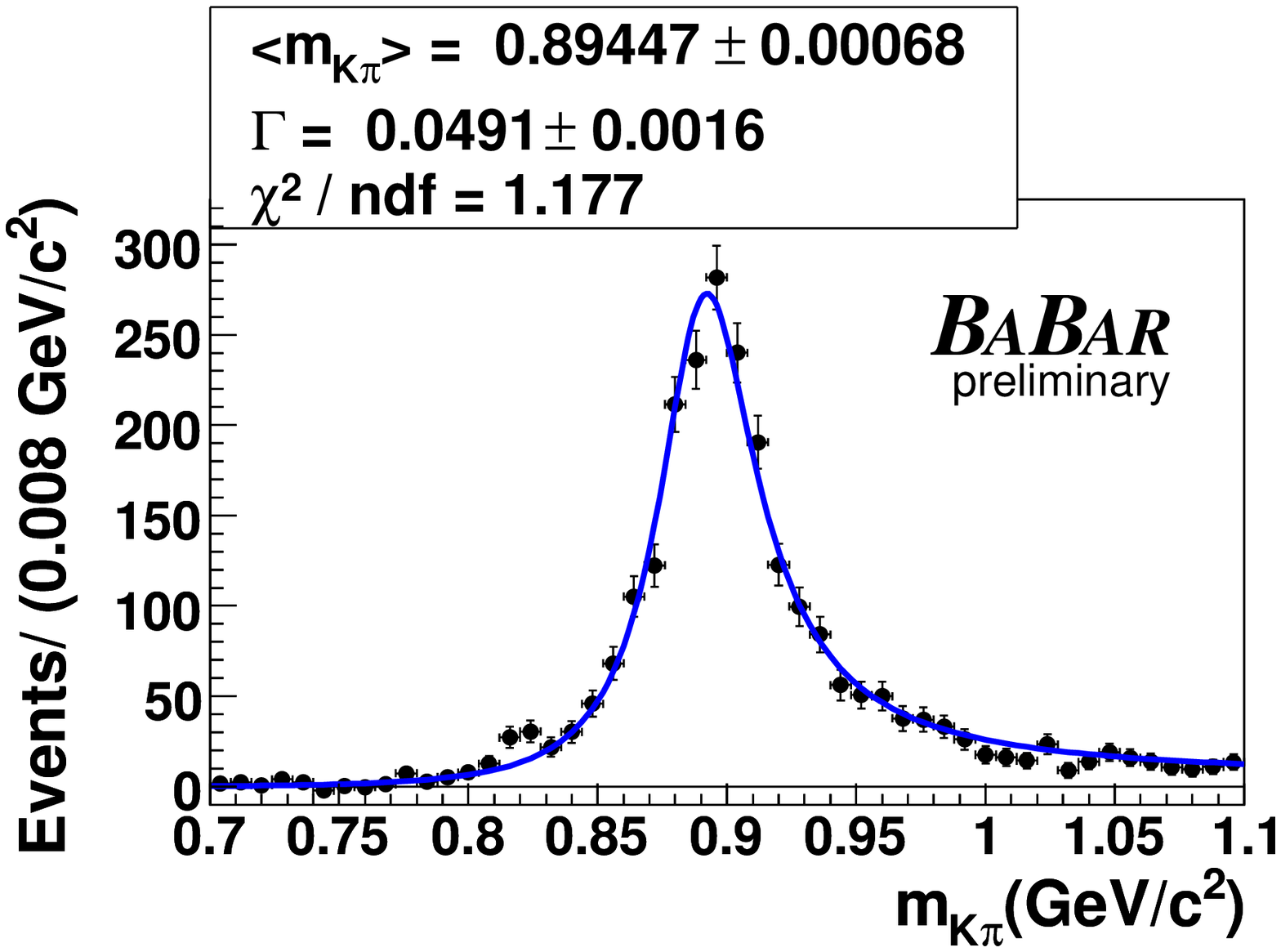}\\
(a) $\kszkp$ mode
        \end{center}
\end{minipage}
\begin{minipage}[t]{0.5\textwidth}
        \begin{center}
        \includegraphics[width = 80mm]{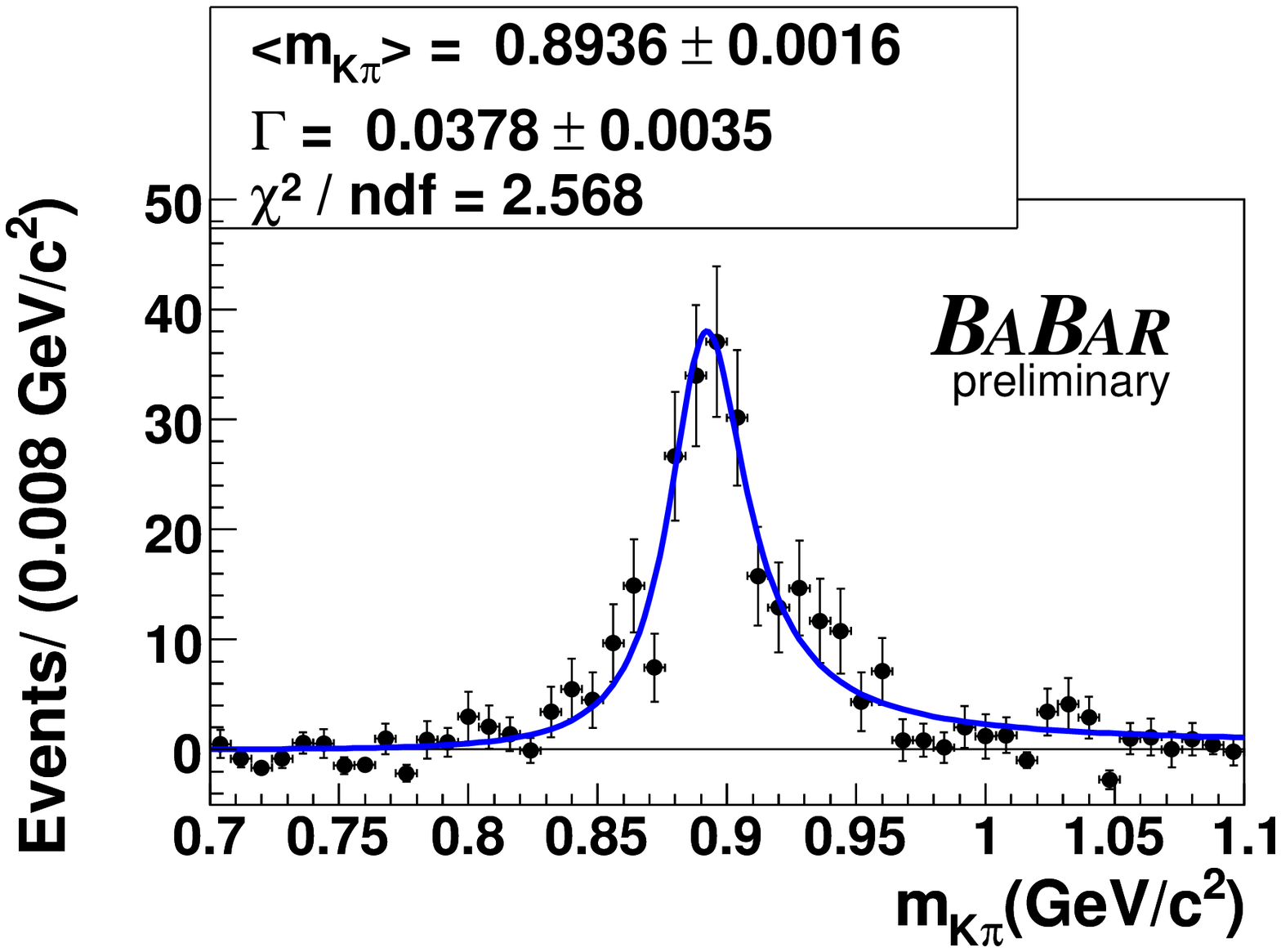}\\
(b) $\kszksz$ mode
        \end{center}
\end{minipage}
\hfill
\begin{minipage}[t]{0.5\textwidth}
        \begin{center}
        \includegraphics[width = 80mm]{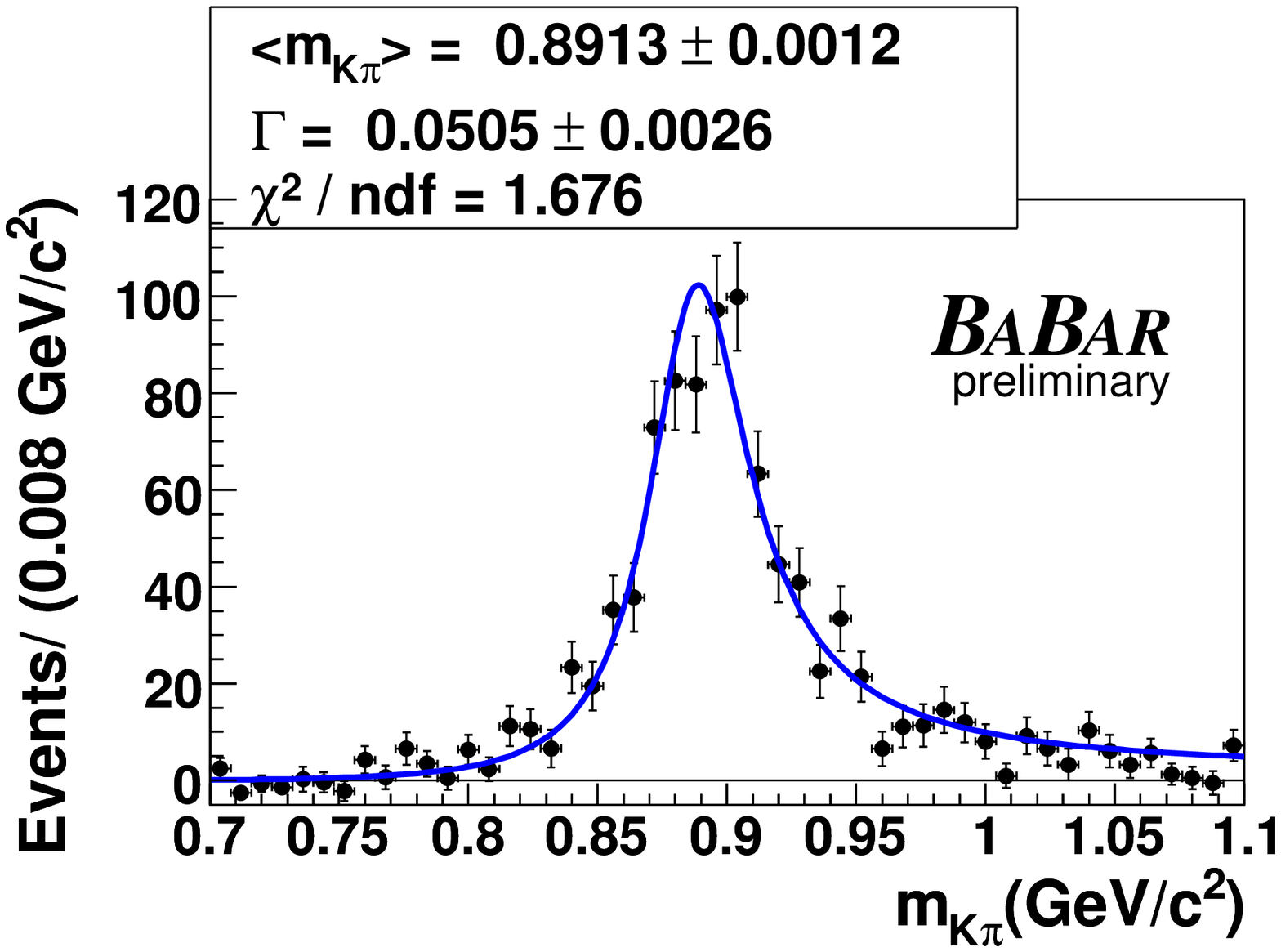}\\
(c) $\kspkpz$ mode
        \end{center}
\end{minipage}
\hfill
\begin{minipage}[t]{0.5\textwidth}
        \begin{center}
        \includegraphics[width = 80mm]{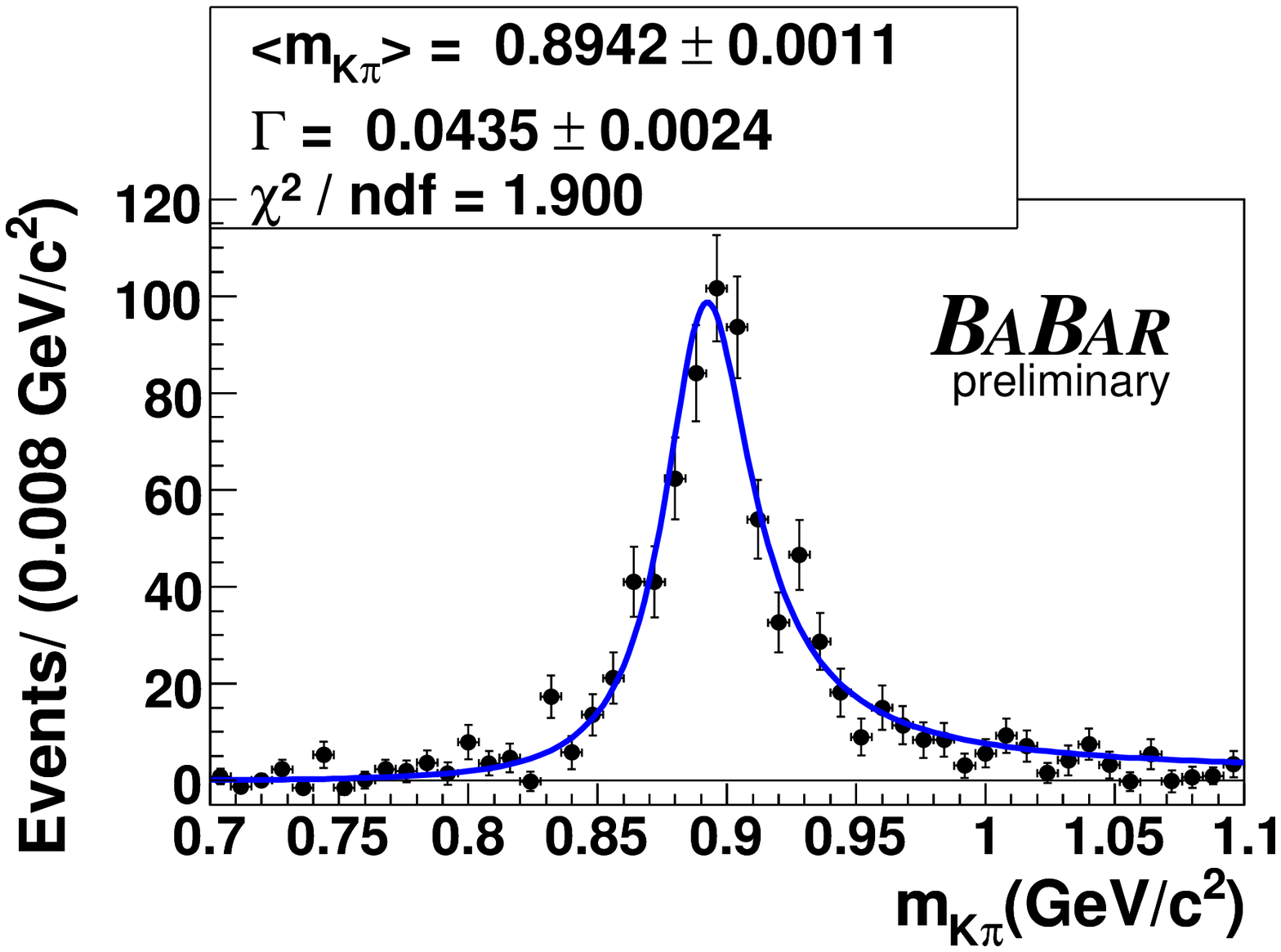}\\
(d) $\kspksp$ mode
        \end{center}
\end{minipage}
\caption{ Relativistic P-wave Breit-Wigner line shape fit to the $K\pi$ invariant mass distribution of the sPlot of data for the 
a) $\kszkp$,~
b) $\kszksz$,~
c) $\kspkpz$,~and
d) $\kspksp$
 modes.}
\label{fig:onpeak_bw}
\end{figure}

To ensure that we are measuring real $K^*$ mesons in data, 
we widen the $K^*$ mass selection to be  
$0.7 < m_{K\pi} < 1.1\GeVcc$, refit the data, and make an 
sPlot~\cite{sPlot} of the $K^*$ mass.  
We then fit a relativistic P-wave Breit-Wigner line shape to the sPlot.   
This is shown in Figure~\ref{fig:onpeak_bw}.  
We combine the measurements of mean of the $K^*$ mass and the 
width for the charged and neutral mesons separately to obtain 
the results in Table~\ref{tab:onpeak_PDG}.  
The results are consistent with the PDG values.
 
\begin{table}[htbp]
        \begin{center}
        \begin{tabular}{c||c|c|c|c}\hline\hline

       &\multicolumn{2}{c|}{Data}&
        \multicolumn{2}{c}{PDG Value}\\\hline

        $K^*$ meson& m (MeV) & $\Gamma$(MeV)& m (MeV) & $\Gamma$(MeV) \\\hline

        $K^{*0}$  & 894.34$\pm~.63$ & 47.1$\pm~1.4$  & 896.00$\pm~.25$ & 50.3$\pm~0.6$ \\  
        $K^{*+}$  & 892.88$\pm~.80$ & 46.7$\pm~1.8$  & 891.66$\pm~.26$ & 50.8$\pm~0.9$ \\\hline\hline

    \end{tabular}
    \end{center}
\caption{The combined results of the fits to the $m_{K\pi}$ spectrum shown in Figure~\ref{fig:onpeak_bw}.  Also shown are the PDG values.}
\label{tab:onpeak_PDG}
\end{table}

\section{CONCLUSIONS}
\label{sec:Conclusions}

We present a preliminary measurement of the branching fractions ${\cal B}(\bkgneut) =  (\kszcombr \pm \kszcomstat \pm \kszcomsyst)\times 
10^{-5}$ and ${\cal B}(\bkpg) =  (\kspcombr \pm \kspcomstat \pm \kspcomsyst) \times
10^{-5}$.  We use these results to calculate the isospin asymmetry at the 90\% confidence interval to be $\isolow < \Delta_{0-} < \isohi.$  
We also present a preliminary measurement of the time-integrated $CP$ asymmetry at the 90\% confidence interval to be $\acplolim < \acp < \acphilim.$  These results are all improvements over previous measurements, as well as being consistent with SM expectations.

\section{ACKNOWLEDGMENTS}
\label{sec:Acknowledgments}

\input pubboard/acknowledgements

\end{document}

%% file: pubboard/authors_ICHEP2008.tex
\begin{center}
\small

The \babar\ Collaboration,
\bigskip

%
B.~Aubert,
M.~Bona,
Y.~Karyotakis,
J.~P.~Lees,
V.~Poireau,
E.~Prencipe,
X.~Prudent,
V.~Tisserand
\inst{Laboratoire de Physique des Particules, IN2P3/CNRS et Universit\'e de Savoie, F-74941 Annecy-Le-Vieux, France }
J.~Garra~Tico,
E.~Grauges
\inst{Universitat de Barcelona, Facultat de Fisica, Departament ECM, E-08028 Barcelona, Spain }
% ...

\end{center}\newpage

%% file: pubboard/acknowledgements.tex
We are grateful for the 
extraordinary contributions of our \pep2\ colleagues in
achieving the excellent luminosity and machine conditions
that have made this work possible.
The success of this project also relies critically on the 
expertise and dedication of the computing organizations that 
support \babar.
The collaborating institutions wish to thank 
SLAC for its support and the kind hospitality extended to them. 
This work is supported by the
US Department of Energy
and National Science Foundation, the
Natural Sciences and Engineering Research Council (Canada),
the Commissariat \`a l'Energie Atomique and
Institut National de Physique Nucl\'eaire et de Physique des Particules
(France), the
Bundesministerium f\"ur Bildung und Forschung and
Deutsche Forschungsgemeinschaft
(Germany), the
Istituto Nazionale di Fisica Nucleare (Italy),
the Foundation for Fundamental Research on Matter (The Netherlands),
the Research Council of Norway, the
Ministry of Education and Science of the Russian Federation, 
Ministerio de Educaci\'on y Ciencia (Spain), and the
Science and Technology Facilities Council (United Kingdom).
Individuals have received support from 
the Marie-Curie IEF program (European Union) and
the A. P. Sloan Foundation.

%% file: paper_la.bbl
\begin{thebibliography}{99}
\bibitem{CC} 
Charge conjugate modes are implied throughout, except for the $CP$ asymmetry.
\end{thebibliography}
